\documentclass[aps,pre,showpacs,reprint,groupedaddress]{revtex4-1}

\usepackage[utf8]{inputenc}
\usepackage{amsmath}
\usepackage{graphicx}
\usepackage{bbm}
\usepackage{url}

\newcommand{\EE}{\mathcal{E}}

\newcommand{\modbp}{mod-bp }
\DeclareMathOperator*{\argmax}{arg\,max}

\begin{document}


\title{Multiple phases in modularity-based community detection}


\author{Christophe Sch\"ulke}
\email{christophe.schulke@espci.fr}
\affiliation{Universit\'e Paris Diderot, Sorbonne Paris Cit\'e, 75205 Paris, France}
\affiliation{Dipartimento di Fisica, Universit\`a di Roma ``La Sapienza'', Piazzale Aldo Moro 2, 00185 Rome, Italy}
\author{Federico Ricci-Tersenghi}
\email{federico.ricci@roma1.infn.it}
\affiliation{Dipartimento di Fisica, INFN--Sezione di Roma 1, and CNR-NANOTEC, UOS di Roma,
Universit\`a di Roma ``La Sapienza'', Piazzale Aldo Moro 2, 00185 Rome, Italy}


\date{\today}

\begin{abstract}
Detecting communities in a network, based only on the adjacency matrix, is a problem of interest to several scientific disciplines.
Recently, Zhang and Moore have introduced an algorithm in [P. Zhang and C. Moore, Proceedings of the National Academy of Sciences {\bf{111}}, 18144 (2014)], 
called \modbp, that avoids overfitting the data by optimizing a weighted average of modularity (a popular goodness-of-fit measure in community detection) and entropy (i.e.\ number of configurations with a given modularity).
The adjustment of the relative weight, the ``temperature'' of the model, is crucial for getting a correct result from \modbp.
In this work we study the many phase transitions that \modbp may undergo by changing the two parameters of the algorithm: the temperature $T$ and the maximum number of groups $q$.
We introduce a new set of order parameters that allow to determine the actual number of groups $\hat{q}$, and we observe on both synthetic and real networks the existence of phases with any $\hat{q} \in \{1,q\}$, which were unknown before.
We discuss how to interpret the results of \modbp and how to make the optimal choice for the problem of detecting significant communities.
\end{abstract}

\pacs{64.60.aq,89.20.-a}

\maketitle

\section{Introduction}
In community detection, the goal is to regroup nodes of an observed network into different groups (or communities)
of nodes believed to be similar, and thus to find a meaningful partition of the network.
The assumption that this is possible comes from the hypothesis that the structure of the graph reflects 
hidden attributes of the nodes, that can therefore be inferred.
Though recent studies show that such an assumption does not hold in general for real networks~\cite{hric2014community},
generative models with this property, such as the stochastic block model~\cite{holland1983stochastic} (SBM), have proved 
the efficiency of community detection algorithms~\cite{decelle2011inference}.
Different classes of community detection algorithm exist: among the popular approaches, 
algorithms relying on Bayesian inference fit the parameters of an assumed generative model to the observed network~\cite{hastings2006community,decelle2011inference}, while 
spectral algorithms find communities from the eigenvectors of a matrix based on the adjacency matrix of the network~\cite{newman2006finding,krzakala2013spectral}.

The hypothesis most commonly made is that of assortative networks, which means that nodes with the same hidden attributes 
are more likely to be linked than nodes with different attributes.
Under this hypothesis of assortative networks, a popular measure of the goodness of a partition is the modularity, and therefore various 
community detection algorithms rely on modularity maximization~\cite{newman2004fast,duch2005community,aloise2010column,cafieri2011locally}.
Recently the authors of \cite{zhang2014scalable} (called ZM hereafter) introduced such an algorithm that avoids the common pitfall 
of overfitting: indeed maximizing modularity predicts communities even in unstructured (i.e.\ random) networks. The only free parameters in the \modbp algorithm are the number of groups $q$ and a temperature-like parameter $T$. 
Three ranges of temperatures are identified that correspond to phases in which the algorithm has qualitatively 
different behaviours: at high temperatures no division in groups is found, at intermediate temperatures meaningful groups are found and for low enough temperatures the algorithm does not converge.
In this paper, we broaden the picture given in ZM by showing that there are in general
more than three phases. We show that despite passing to the \modbp algorithm the number $q$ of groups, it can spontaneously return a partition with a smaller number of groups $\hat{q}<q$. 
We introduce a new set of order parameters that allows to determine $\hat{q}$, and observe both on synthetic and real networks the existence of phases with different values of $\hat{q}\in\{1,q\}$.


We will use the following notations: $N$ is the number of nodes in the network, $\EE$ is the set of $m$ undirected edges, and we write 
$\langle i j \rangle \in \EE$ if an edge is present between nodes $i$ and $j$. The degree $d_i$ of a node
is the number of edges that link node $i$ to other nodes. A partition of the network is a set $\{t\}$, where $t_i\in\{1,q\}$ is the group node $i$ belongs to. $q$ is the maximum number of groups.

The modularity of a partition $\{ t \}$ is defined by \cite{newman2004finding}
\begin{equation}
 Q(\{t\}) = \frac{1}{m} \left( \sum_{\langle i j \rangle \in \mathcal{E}} \delta_{t_i,t_j} - \sum_{\langle i j \rangle} \frac{d_i d_j}{2 m} \delta_{t_i,t_j} \right) \;,
\label{eq:modularity}
\end{equation}
where $\delta$ is the Kronecker delta function.
High values of modularity indicate that there are more edges between nodes of the same group 
than between nodes of different groups: thus, the higher the modularity, the better the partition.
The advantage of modularity is that it makes no assumption on the way the network was generated, but 
only that it has an assortative structure. 
This encourages its use on real networks, in which the true generative process is generally unknown, and 
has lead to several algorithms performing community detection by maximization of modularity~\cite{newman2004fast,duch2005community,aloise2010column,cafieri2011locally}.

One drawback of modularity is that finding the partition with highest 
modularity is a discrete combinatorial optimization problem~\cite{brandes2008modularity}, which becomes rapidly intractable as $N$ increases;
so effective heuristics have to be developed. Another drawback is that modularity maximization is prone to overfitting: 
it is possible to find high-modularity partitions even in Erd\H{o}s-Renyi random graphs ~\cite{erdHos1959random} 
although by construction they do not contain an underlying group strucure~\cite{guimera2004modularity,reichardt2006statistical,lancichinetti2010statistical}. 
Finally, there exists a fundamental resolution limit~\cite{fortunato2007resolution}, that prevents from recovering small-sized groups.

ZM introduces a new community-detection algorithm based on modularity maximization, 
tackling the two first mentioned drawbacks, and proposing a multi-resolution strategy to overcome the third. The algorithm, called \modbp, is scalable, i.e. is of polynomial complexity with 
respect to $N$, and the authors show that it does not overfit, in the sense that it does not return high-modularity partitions 
for Erd\H{o}s-Renyi networks.

This is achieved by treating modularity maximization as a statistical physics problem with an energy
\begin{equation}
 E(\{t\}) = -m Q(\{t\}) \label{eq:energy}
\end{equation}
at a finite temperature $T=\frac{1}{\beta}$. In this way, every partition $\{ t\}$ is given 
a probability taken from the Gibbs distribution
\begin{equation}
 P(\{t\}) = \frac{1}{Z} e^{- \beta E(\{t \})} ,
 \label{eq:distribution}
\end{equation}
where $Z$ is the partition function
\begin{equation}
 Z = \sum_{\{t\}} e^{-\beta E(\{t\})}.
\end{equation}

To solve the problem of sampling from the Gibbs distribution~(\ref{eq:distribution}), ZM proposes a belief propagation (BP) algorithm~\cite{MezardMontanari09, yedidia2003understanding}, in which so-called messages 
$\psi_t^{i \to k}$ are sent between all pairs of nodes $ \langle i k \rangle$, for $q$ different groups $t$.
We refer the reader to ZM for a precise description of the algorithm. 
After convergence of the BP algorithm, marginals $\psi_t^i$ are obtained from the messages. $\psi_t^i$ represents the probability 
that node $i$ belongs to group $t$, and the most-likely group for node $i$ is therefore:
\begin{equation}
 \hat{t}_i = \argmax_t \psi_t^i .
 \label{eq:argmax}
\end{equation}
Using this maximization, the maximum a posteriori modularity $Q^{\rm MAP}$ 
corresponding to the assignment $\{\hat{t} \}$ can be calculated as
\begin{equation}
 Q^{\rm MAP} = Q(\{\hat{t}\}).
\end{equation}

As the algorithm samples from the distribution (\ref{eq:distribution}), one can also define an average modularity 
$Q^{\rm MARG}$ that is calculated from the marginals instead of the most-likely partition, 
and which is proportional to the average energy of the model
 \small
 \begin{equation}
 Q^{\rm MARG } = \frac{1}{m} \frac{\partial}{\partial \beta} \left( \sum_i \log Z_i -\sum_{\langle i j \rangle \in \mathcal{E}} \log Z_{ij} +  \frac{\beta}{4m} \sum_t \theta^2_t \right), \nonumber
 \label{eq:Qmarg}
 \end{equation}
 \normalsize
where $Z_i$ and $Z_{ij}$ are the normalizations of the marginals and the two-point correlation functions respectively, and $\theta_t = \sum_{j=1}^N d_j \psi_t^j$.

While the problem of maximizing modularity is equivalent to finding the 
ground state of (\ref{eq:energy}),
sampling from (\ref{eq:distribution}) at a finite temperature corresponds to minimizing the corresponding free energy.
This means taking into account not only the modularity, but also the entropy, counting the number of partitions with a given modularity. 
In this way, instead on focussing on a single partition, \modbp at finite $T$ returns a partition that is a good consensus 
of the many existing high-modularity partitions, as advocated in~\cite{lancichinetti2012consensus}.

\section{Phase transitions}

As in numerous statistical physics problems, (\ref{eq:distribution}) may lead to phase transitions at some given temperatures.
Using modularity as an energy function is similar to study a Potts-like statistical mechanics problem~\cite{hu2012phase}, for which Ref.~\cite{ronhovde2012global} has shown that a phase transition is always present.
ZM reports that temperature ranges define three different regimes of the algorithm. At very 
low temperatures, the system is in a spin glass phase, in which the algorithm does not converge to a fixed point.
At high temperature, the system is in a paramagnetic phase in which the fixed point is trivial and  all nodes have an equal probability $1/q$ of belonging to 
any of the groups. In networks with statistically significant communities, there is an intermediate temperature range called 
recovery phase, in which the algorithm converges to a non-trivial fixed point, from which group assignments can be obtained 
using (\ref{eq:argmax}).

Here, we broaden this picture by showing that the recovery phase itself can be divided in up to $q-1$ phases, with $2\le\hat{q}\le q$. 
At a temperature separating two phases, there is an order parameter that becomes vanishingly small, increasing $T$, and the number of iterations 
needed by the algorithm to reach the fixed point diverges.

\subsection{Model-based critical temperatures}
Modularity as a measure of goodness of a partition is particularity appealing for real networks, because it makes no assumption 
about an underlying model that generates the network. 
Though appealing, this absence of model is problematic when it comes to determining the best temperature at which to run \modbp (i.e.\ there is not Bayes optimal temperature).
In ZM two generative models are analyzed, allowing to find two useful characteristic temperatures:
\begin{enumerate}
 \item In the configuration model, a network is built by randomly creating links between nodes of known degree, 
 until all nodes have the right number of neighbours.
ZM shows that in this model, the phase transition between the spin-glass and 
 the paramagnetic phase takes place at
 \begin{equation}
  T^* = \left( \log \left( \frac{q}{\sqrt{c}-1} +1 \right) \right)^{-1}, 
  \label{eq:tstar}
 \end{equation}
 where $c$ is the average excess degree, calculated from the average degree $\langle d \rangle$ and 
 the average squared degree $\langle d^2 \rangle$, and given by
 \begin{equation}
  c=\frac{\langle d^2 \rangle}{\langle d \rangle} -1 .
  \label{eq:c}
 \end{equation}
 
 \item In the stochastic block model (SBM) \cite{holland1983stochastic}, the nodes are grouped into $q^*$ equal-sized groups,
 and for each pair of nodes $ \langle i j \rangle$, a link is created with probability $p_{rs}$ if $i$ belongs to 
 group $r$ and and $j$ belongs to group $s$. In the most simple case, we take $p_{rs}=p_{\rm out}$ if $r \neq s$ 
 and $p_{rs}=p_{\rm in}$ if $r=s$. One often condiders networks with sparse connectivity, i.e. the average 
 number of links between a node $i$ from group $r$ and all the nodes from group $s$, $c_{rs}$, does not grow with the size 
 of the network.
 ZM shows that \modbp is as successful as a Bayes-optimal algorithm 
 \cite{decelle2011asymptotic} and that the transition between the paramagnetic phase and the recovery phase takes place at
 \begin{equation}
  T_R(\epsilon) = \left( \log \left( \frac{q (1 + (q-1) \epsilon)}{c(1-\epsilon) - (1+(q-1) \epsilon)} +1 \right) \right)^{-1} ,
  \label{eq:tr}
 \end{equation}
 where $\epsilon = p_{\rm out}/p_{\rm in}$.

\end{enumerate}

For real retworks, the stochastic block model is
usually a bad model, and the recommendation of ZM is to run the algorithm at $T^*$, which seems 
to always lie inside of the recovery phase.
We can also note that the $\epsilon \to 0$ limit of (\ref{eq:tr}),
\begin{equation}
 T_0 = \left( \log \left( \frac{q}{c-1} +1 \right) \right)^{-1}
\end{equation}
is a useful upper bound for $T$. Indeed, above this temperature, the algorithm converges to the paramagnetic solution even 
for networks composed of disconnected components, and is therefore useless. 

\subsection{Degenerate groups}
In the paramagnetic phase, the marginal $\psi_t^i$ of every node $i$ of the network is equal to $\frac{1}{q}$ for all $t$, 
up to some minor fluctations due to the numerical precision of the machine or incomplete convergence of the algorithm. 
Due to those fluctuations, calculating a retrieval configuration with (\ref{eq:argmax}) is in general still possible, 
and would lead to a very small but non-vanishing retrieval modularity $Q^{\rm MAP}$.

However, the meaning of the paramagnetic phase is that all groups are strictly equivalent or degenerate, and therefore 
$Q^{\rm MAP}$ should be exactly zero. In order to obtain this, the algorithm has to check for degenerate groups before 
assigning a group to each node, and assign the same ``effective'' group to nodes for which the maximization (\ref{eq:argmax}) leads to 
different, but actually degenerate groups.

To check if groups are degenerate, we can look at the following distance between two groups $k$ and $l$:
\begin{equation}
 d_{kl} = \frac{1}{N} \sum_{i=1}^{N} ( \psi^i_k - \psi^i_l)^2.
 \label{dkl}
\end{equation}
If $d_{kl}$ is smaller than a choosen threshold $d_{\rm min}$, we can consider that group $k$ and group $l$ are degenerate 
and that they should not be distinguished.

\begin{figure}
\centering
 \includegraphics[width=0.9\columnwidth]{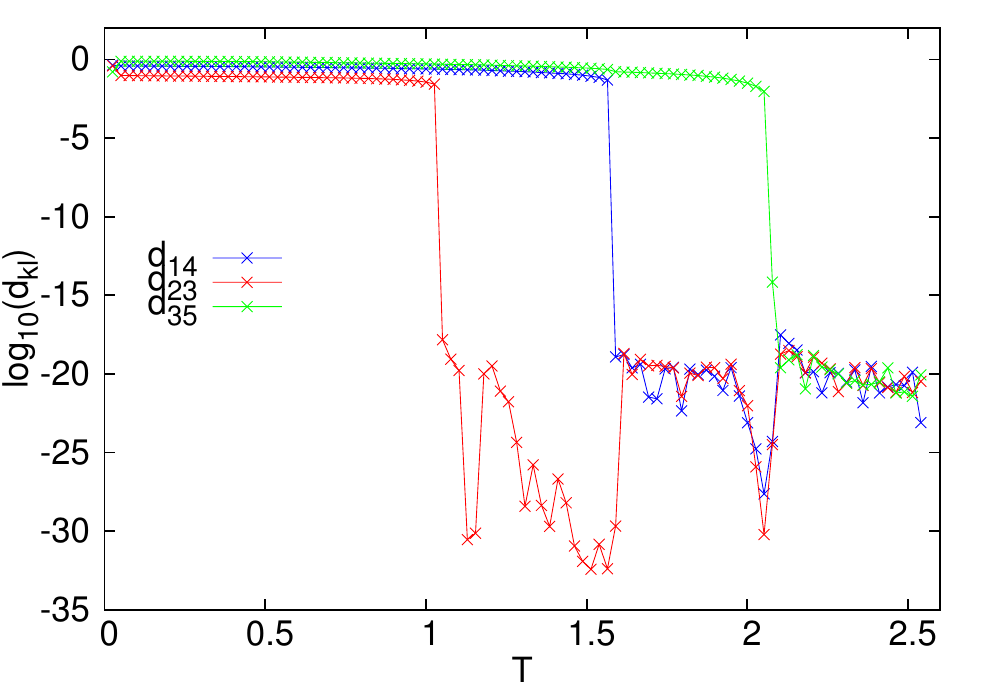}
 \caption{(Color online) $d_{12}$, $d_{23}$ and $d_{35}$ as a function of temperature. In order to follow the groups at different temperatures, 
the temperature is increased step by step, and the messages are initialized with the final values they reached at the last temperature. We see that the group distances $d_{kl}$ are like order parameters undergoing a phase 
transition at different temperatures, where they drop by more than ten orders of magnitude.
Due to this phase transition, it is easy to choose a threshold $d_{\rm min}$ in Eq.~(\ref{eq:phi}).
The dataset is ``political books'', run with $q=6$.}
\label{fig:1}
\end{figure}

\begin{figure*}[floatfix]
\centering
\includegraphics[width=0.33\columnwidth]{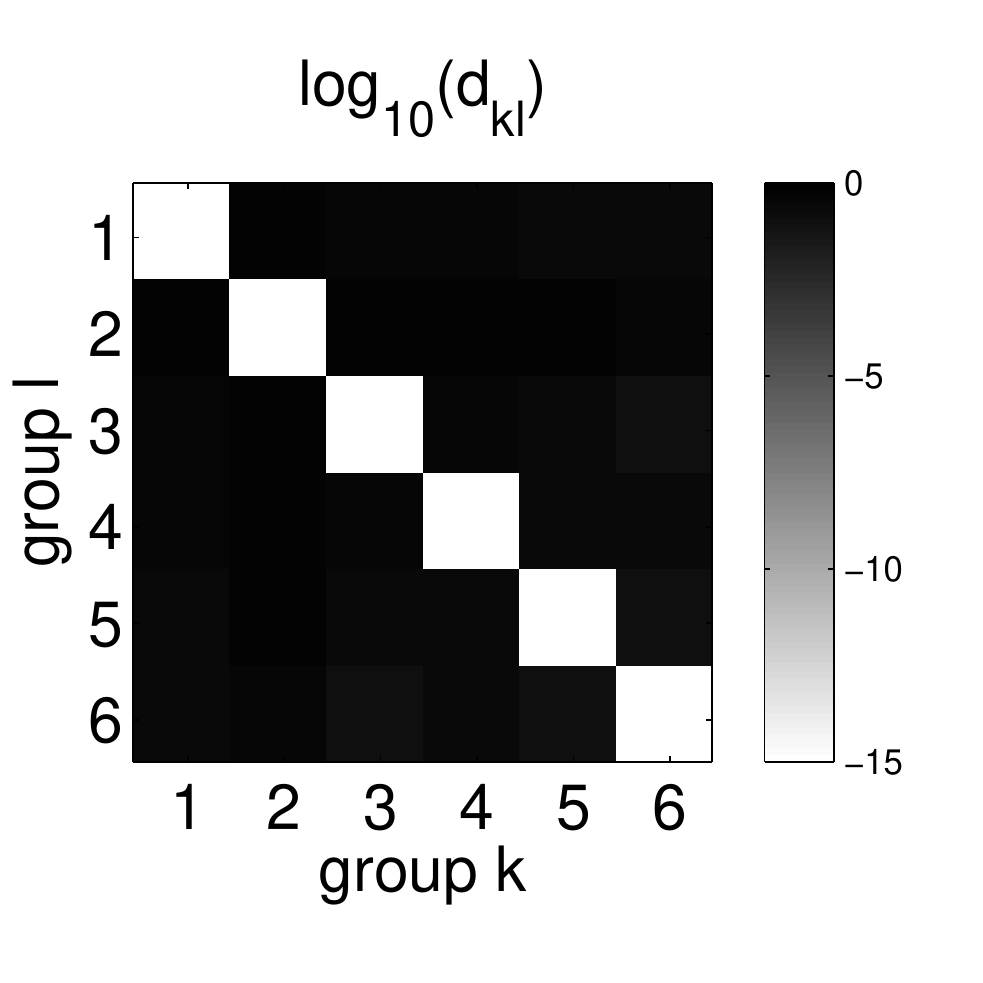}
\includegraphics[width=0.33\columnwidth]{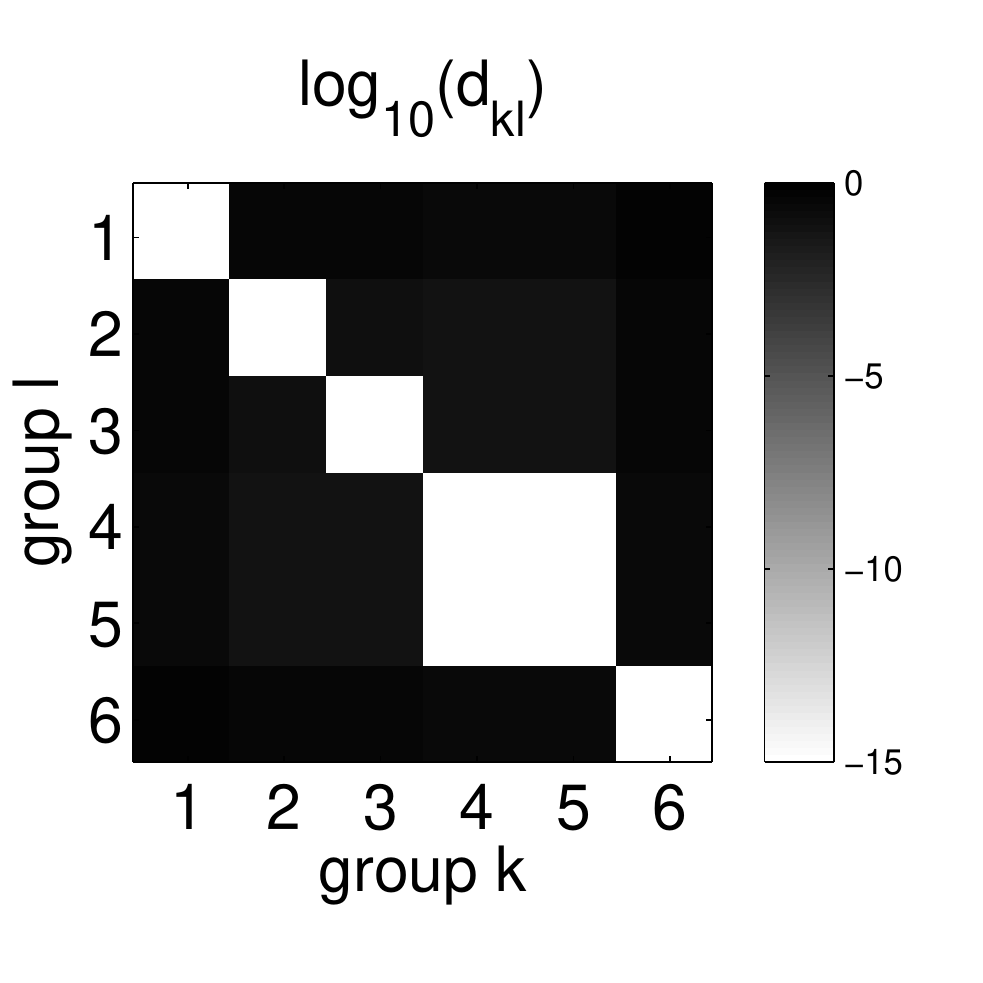}
\includegraphics[width=0.33\columnwidth]{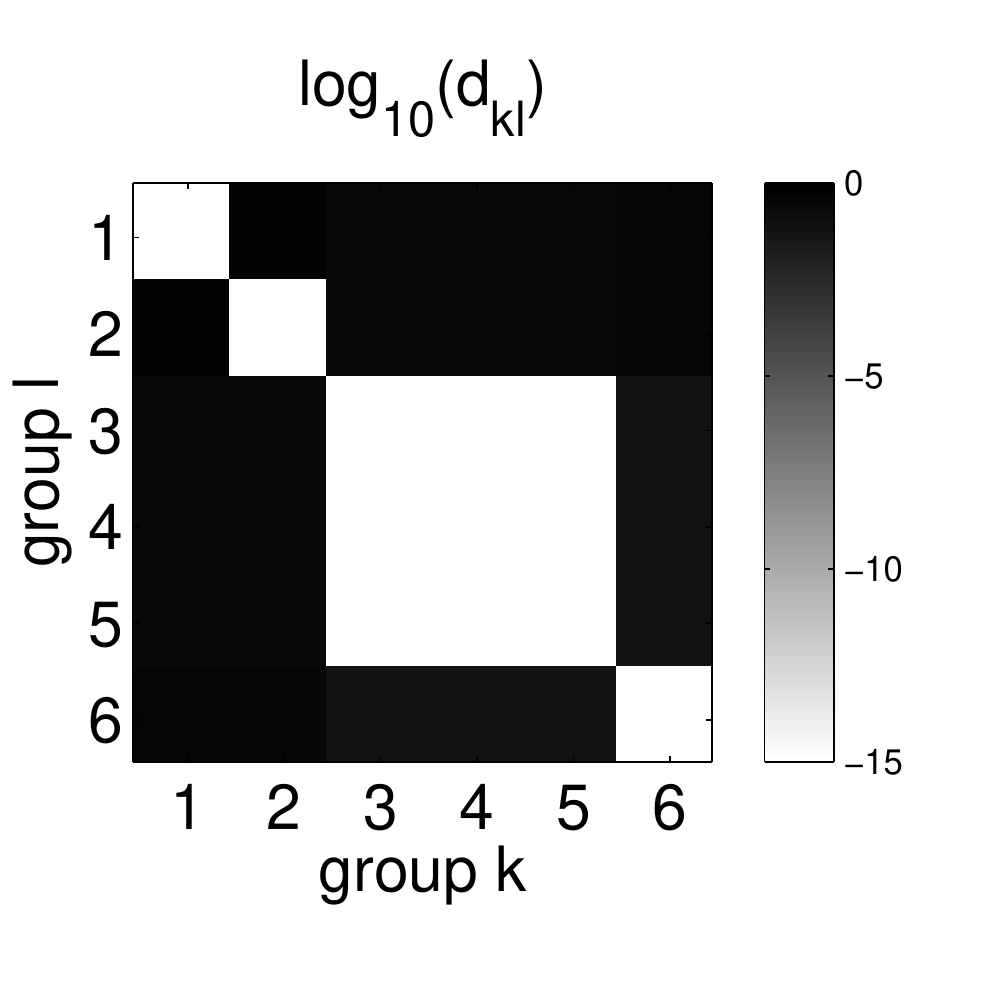}
\includegraphics[width=0.33\columnwidth]{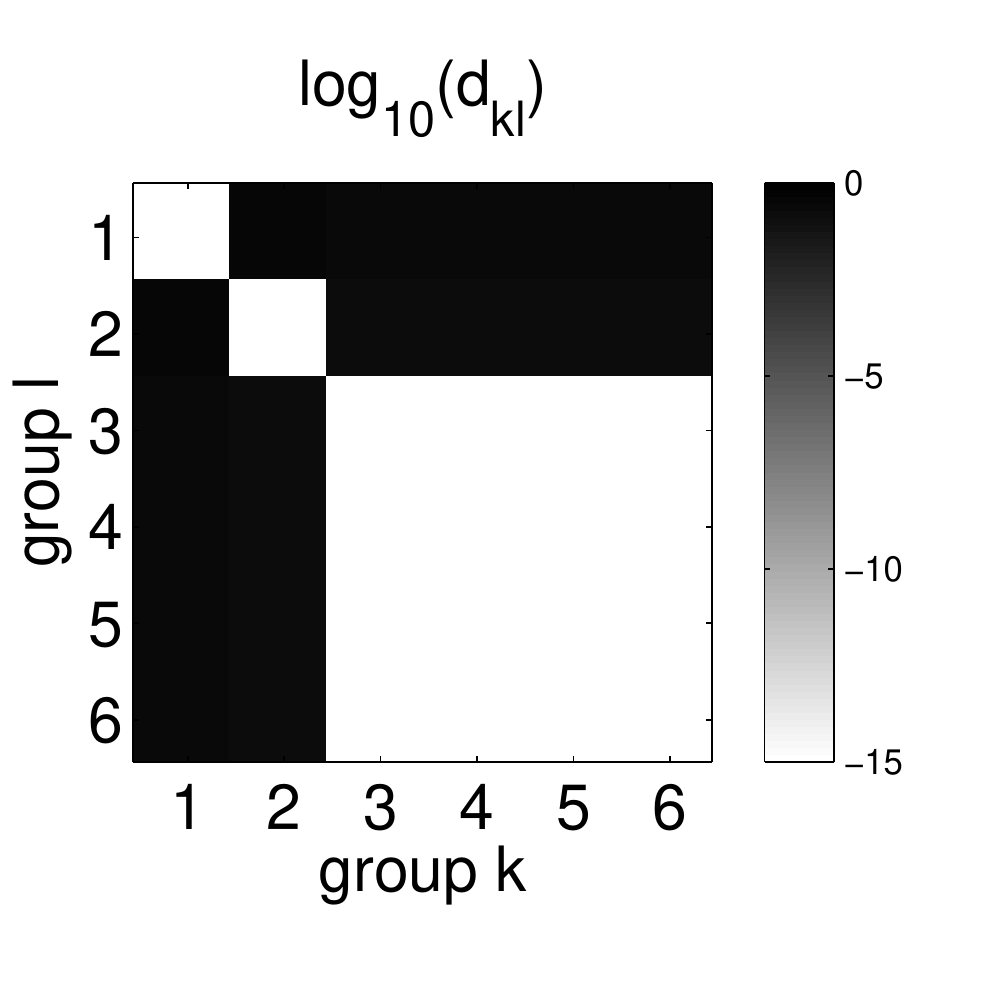}
\includegraphics[width=0.33\columnwidth]{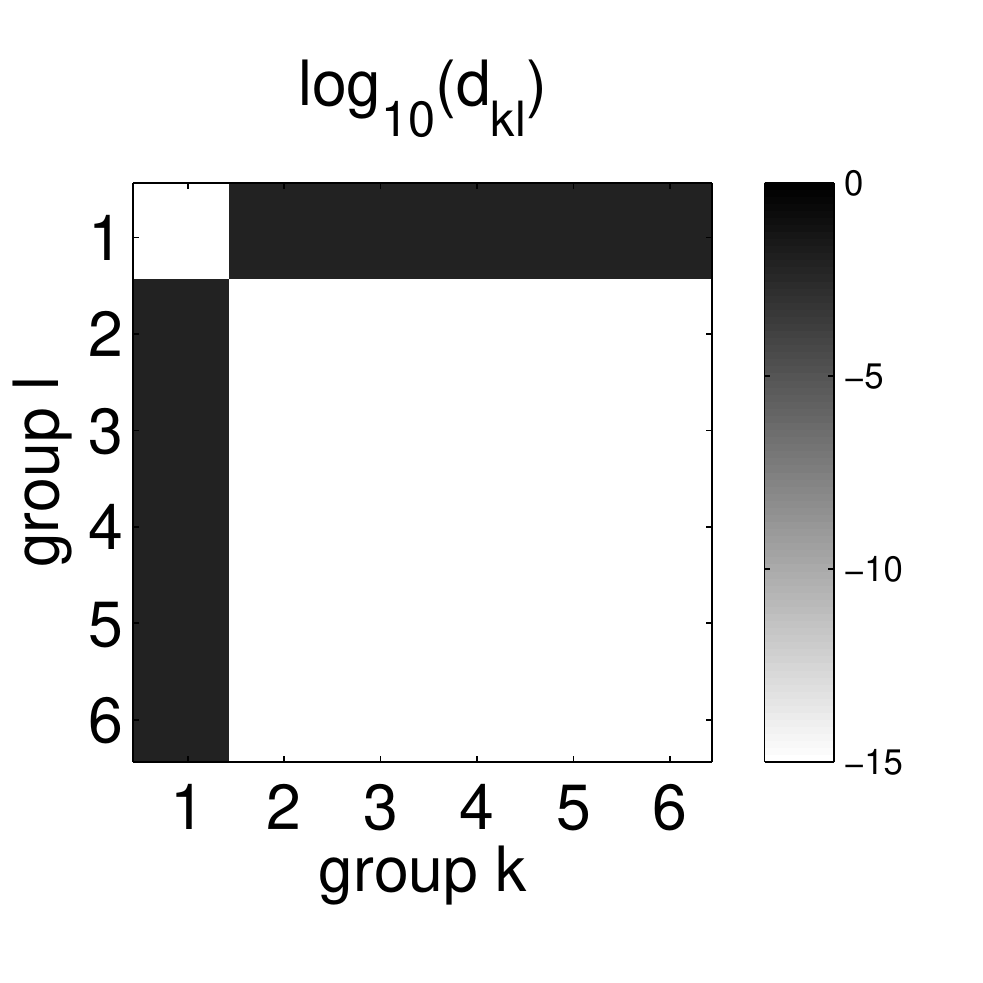}
\includegraphics[width=0.33\columnwidth]{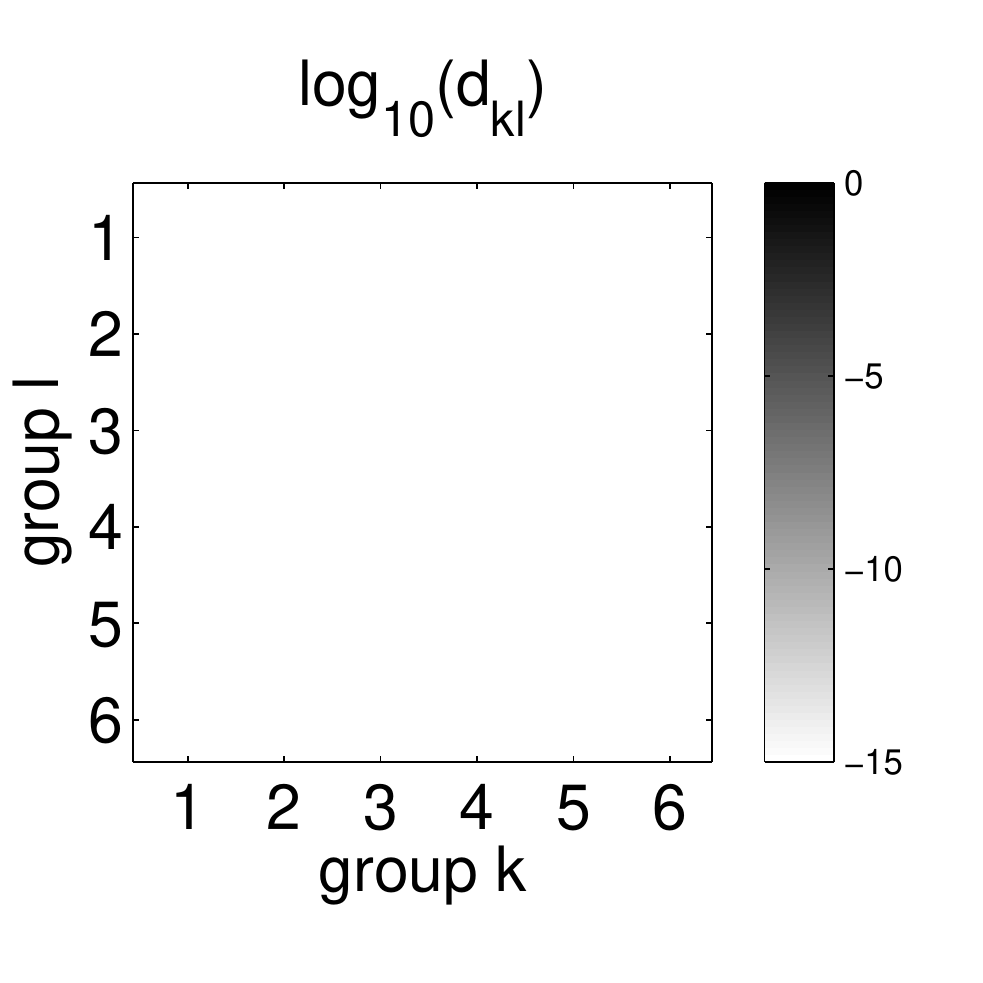}
\caption{Matrices of distances between groups for different temperatures. The dataset is ``political books'', the algorithm is run with $q=6$ for $T= 0.26, 0.9, 1.0, 1.28 , 2.1, 2.3 $ (from left to right). We observe the formation of a growing cluster of groups that are equivalent, allowing us to define a number of effective groups $\hat{q}$, that varies from $6$ at low temperature (left) to $1$ in the paramagnetic phase (right). Note that the area of the squares is not related to the number of nodes contained in the groups.}
\label{fig:2}
\end{figure*}

An effective number of groups, $\hat{q}$, can then be defined as the number of \emph{distinguishable} groups.
We can define a mapping $\phi$ between the $q$ groups used by the algorithm and the $\hat{q}$ distinguishable groups: 
for each group $k$, $\phi(k)$ is an integer between $1$ and $\hat{q}$ representing one of the effective groups, and
\begin{equation}
 \forall (k,l), \quad \phi(k) = \phi(l) \Leftrightarrow d_{kl}<d_{\rm min}.
 \label{eq:phi}
\end{equation}
With this mapping, we replace the group assignment procedure (\ref{eq:argmax}) by 
\begin{equation}
 \hat{t}_i = \phi \left( \argmax_t \psi_t^i \right).
 \label{eq:assignment}
\end{equation}
With this assignment procedure, $Q^{\rm MAP}$ is strictly zero in the paramagnetic phase, because all nodes belong to the same group. 

Fig.~\ref{fig:1} shows that choosing a threshold $d_{\rm min}$ is meaningful because $d_{kl}$ undergoes a phase transition at which it sharply drops of several orders of magnitude.

Interestingly, group degeneracy is not only observed in the paramagnetic phase, but also inside the retrieval phase. 
In that case, not all groups are degenerate, but only a subset of them. Fig.~\ref{fig:2} shows this for the popular network 
``political books''~\cite{polbooks}, on which \modbp was run at different temperatures.

\section{Existing domains of phases}
 
\begin{figure}
 \centering
 \includegraphics[width=0.48\textwidth]{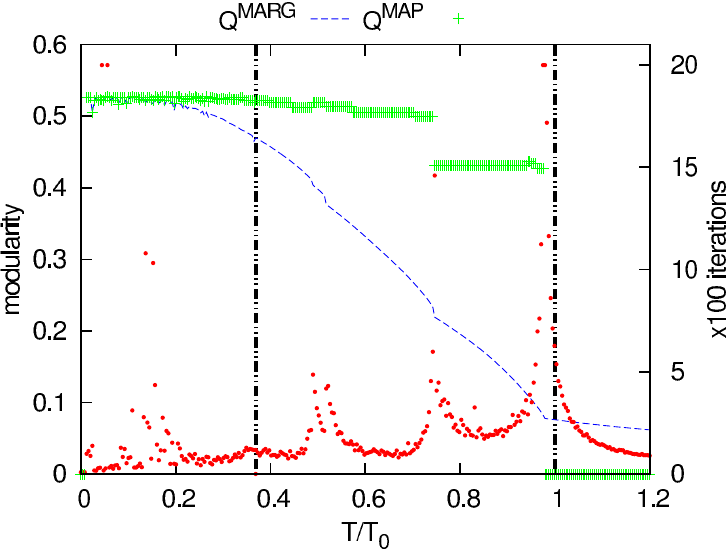}
 \includegraphics[width=0.48\textwidth]{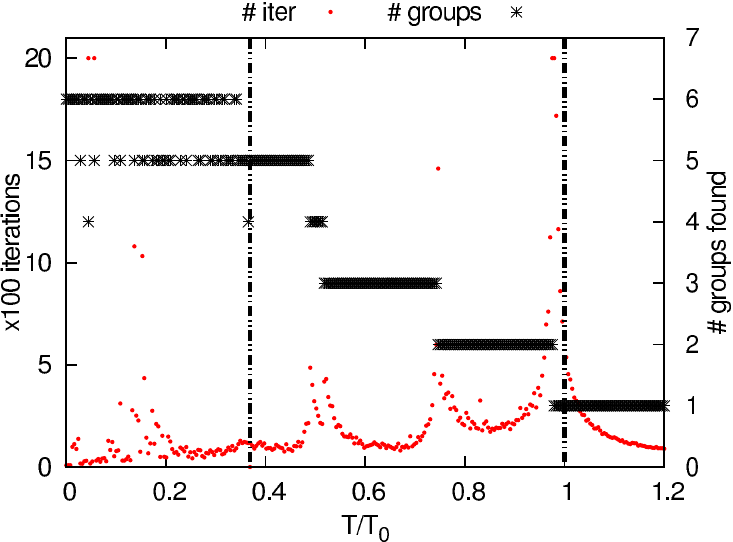}
 \caption{(Color online) Modularities and numbers of effective groups $\hat{q}$ obtained by sweeping a temperature range from $0$ to $1.2$  $T_0$ on the dataset ``political books'' with $q=6$.
The vertical lines indicate the positions of $T^*$ (left) and $T_0$ (right). Above $T^*$, the changes in $\hat{q}$ define quite homogeneous phases, separated by sharp transitions, where the number of iterations necessary to reach convergence increases greatly. Below $T/T_0 \approx 0.4$, the phase is not homogeneous: depending on the starting conditions, $\hat{q}$ can be $4$, $5$ or $6$. Note that $Q^{\rm MAP}$ increases only minimally when $\hat{q}$ exceeds $3$, which agrees with the fact that $q^*=3$.}
\label{fig:3}
\end{figure}

Thanks to the group assignment procedure in Eq.~(\ref{eq:assignment}), up to $q+1$ phases can exist for any network on which \modbp is run with $q$ groups: one for each $\hat{q}\in\{1,q\}$, plus a spin glass phase.
Fig.~\ref{fig:3} shows this for the network ``political books''.
On this network, several phases coexist at low temperature, whereas for higher temperatures, the phases exist in well separated temperature intervals.  In the latter case, we can define a `critical' temperature $T_k$, separating the phase with $\hat{q}=k$ from the one with $\hat{q}=k+1$.
As can be seen on Fig.~\ref{fig:3}, the number of iterations needed for \modbp to converge greatly increases around these critical temperatures.
As noted before, $T_0$ is a good reference temperature, and normalizing all temperatures by $T_0$ is a good way to compare 
critical temperatures $T_k$ for the same network with different $q$ values, and also for comparing different networks.

\subsection{Location of critical points $T_k$}

\begin{figure}
 \centering
 \includegraphics[width=0.45\textwidth]{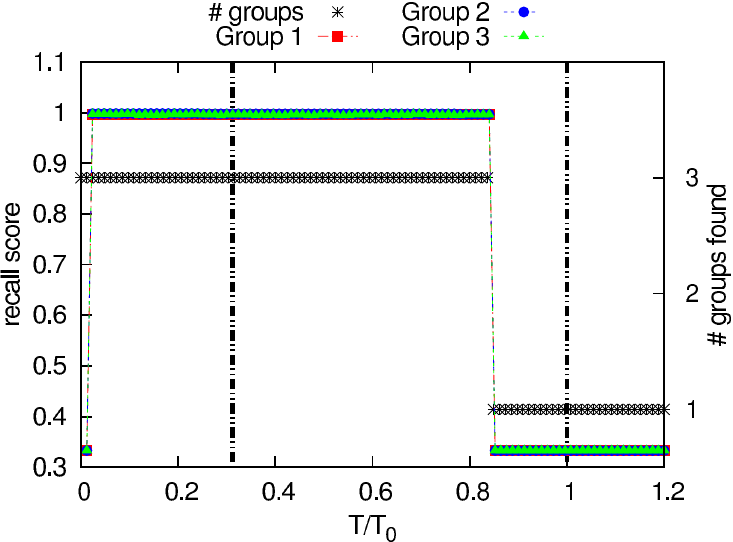}
 \includegraphics[width=0.45\textwidth]{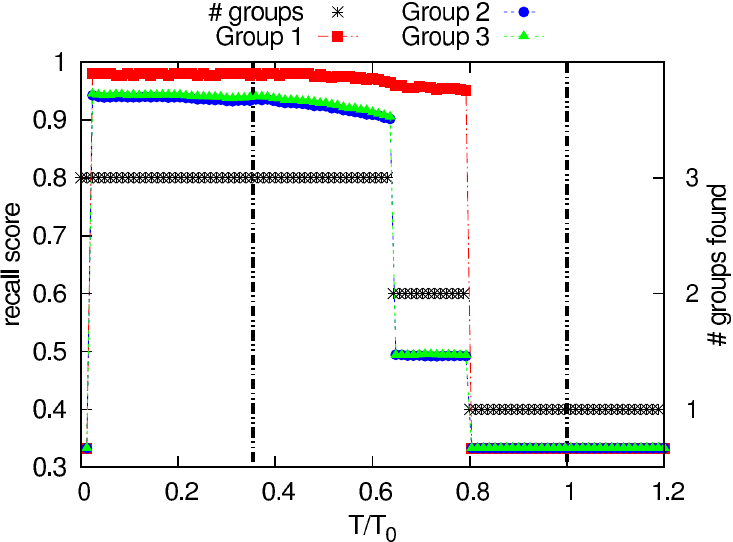}
 \caption{(Color online) Degeneracy of $T_k$s on networks generated by the SBM with $N=5000$, $q^*=3$ and $c_{\rm out}=2$. \modbp was run with $q=3$.
 \textbf{Top:} group 1 has higher in-connectivity than the two others: $c_{11}=30$ whereas $c_{22}=c_{33}=15$. 
 $T_1$ and $T_2$ are distinct, and from the recall 
 scores we see that only group 1 is detected between $T_1$ and $T_2$, whereas groups 2 and 3 have an equally low recall score, 
 as in the partition given by the algorithm, they are merged to a single group. Below $T_2$, $\hat{q}=3$ and the algorithm 
 separates groups 2 and 3.
 \textbf{Bottom:} all 3 groups have the same in-connectivity $c_{\rm in}=30$. There is no $\hat{q}=2$ phase because $T_1$ and $T_2$ are degenerate. 
 For both experiments, the spin-glass phase is not reached.}
 \label{fig:4}
\end{figure}

In some cases, a subset of $n$ critical temperatures $T_k$ can be degenerate, in which case there is a phase transition 
between a phase with $\hat{q}=k$ and a phase with $\hat{q}=k+n$. For instance, this is the case in networks generated by 
the stochastic block model with the same in-connectivity $p_{in}$ inside each of the $q^*$ groups (Fig.~\ref{fig:4}, above).
This agrees with the description of the three phases given in ZM. 

In contrast, in networks generated 
by the SBM with $p_{rr} \neq p_{tt}$ if $r\neq t$, the degeneracy is lifted (Fig.~\ref{fig:4}, below).
The figure also shows that, starting above $T_0$ (i.e. in the paramagnetic phase) and lowering the temperature, 
the groups are inferred in order of their strength.

To show this, we use the recall score for different groups, which allows to see if one of the inferred 
groups corresponds well to a given real group. To quantify the similarity between a real group $G$ and 
an inferred group $\hat{G}_i$, that are not necessarily of the same size, we can use the Jaccard score \cite{hric2014community}, which is defined by:
\begin{equation}
 J(G,\hat{G}_i)= \frac{|G \cap \hat{G}_i|}{|G \cup \hat{G}_i|}.
\end{equation}
The recall score is the maximum of the Jaccard score:
\begin{equation}
 R(G) = \max_{i} J(G,\hat{G}_i).
\end{equation}
A recall score close to $1$ means that one of the inferred groups $\hat{G}_i$ is almost identical to group $G$.
Fig.~\ref{fig:4} (below) therefore shows that around $T/T_0=0.7$, the group with the biggest in-connectivity is 
nearly exactly recovered by one of the groups returned by the algorithm, whereas the two groups with lower in-connectivity
 are not. Only by further lowering temperature, when $\hat{q}=3$, all the groups are correctly inferred.

\subsection{Running \modbp with $q \neq q^*$}
On networks generated with the SBM, the real number of groups $q^*$ is known, and it is thus interesting 
to look at what happens when \modbp is run with $q \neq q^*$. 
The behaviour for $q=q^*$ is described in ZM and in Fig.~\ref{fig:4}.
If $q<q^*$, \modbp cannot return the right number of groups, and will merge some of the real groups together so as to obtain $q$ groups.
The more insteresting case  is when $q$ is bigger than $q^*$.

First of all, the range of temperatures of the spin glass phase grows as $q$ increases.
If $\epsilon$ is only slightly above the detectability threshold $\epsilon^*$ \cite{decelle2011asymptotic,mossel2012stochastic}, increasing $q$ can lead to a situation where there is no recovery phase, between the paramagnetic phase and the spin-glass phase. 

However, we will focus on the case when $\epsilon$ is small enough for intermediate phases to be present.
  As described previously, the phase transitions are degenerate if $p_{\rm in}$ is the same for all groups. 
  Therefore, we generally observe only one intermediate phase, with $\hat{q}=q^*$. However, this is not always the case and 
  \modbp can return partitions with different $\hat{q}$ values, depending on the intialization, similarily to what is observed on the 
  real network in Fig.~\ref{fig:3}. 
  Two phenomena can be observed, separately or simultaneously. 
  
\begin{figure*}
\centering
\includegraphics[width=0.5\textwidth]{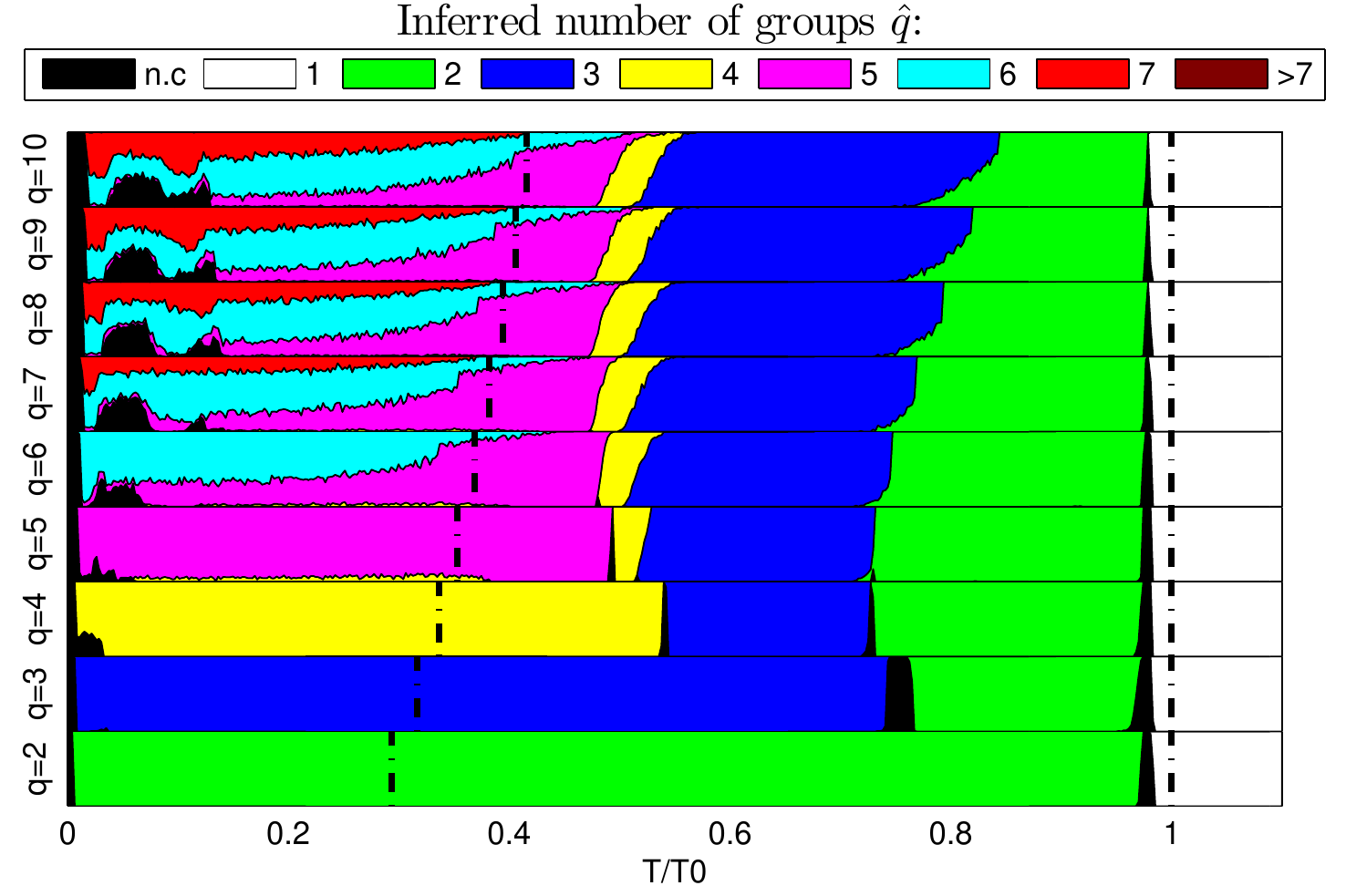}
\includegraphics[width=0.49\textwidth]{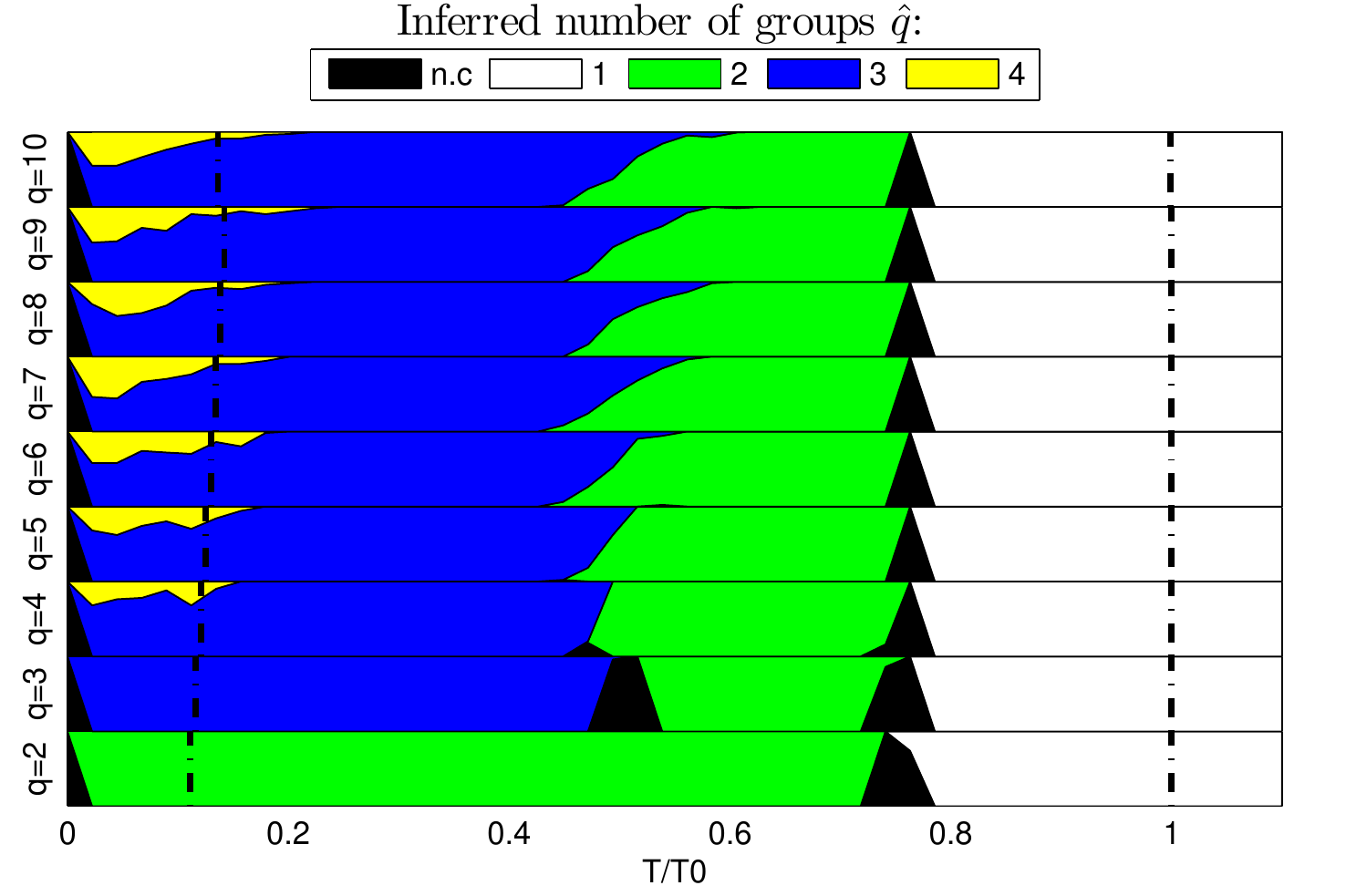}
\caption{(Color online) These plots show the inferred number of groups $\hat{q}$ as a function of the normalized temperature $T/T_0$ and of $q$, for the ``political books'' (left) and ``political blogs'' (right) networks. The dotted lines mark $T=T^*$ (left) and $T=T_0$ (right). 
The ``n.c'' areas correspond to instances that did not reach the convergence criterion ($10^{-6}$) in 700 and 300 iterations, respectively for the two networks.
To take into account coexisting phases, the algorithm was run for $200$ (respectively $50$) different initializations at each temperature.
The position of $T_1$ is very stable across the different values of $q$, and is characterized by a diverging number of iterations.
The other critical temperatures $T_k$ are not always well defined due to overlaps between phases, and to phase transitions becoming much less sharp; however, up to $q=4$, the phases stay well separated, with a clear divergence of the number of iterations.
Remarkably, the existence domains of each phase in terms of $T/T_0$ does not vary a lot with $q$.
 }
 \label{fig:7}
\end{figure*}

  The first phenomenon is the one with $\hat{q}=q^*+1$, where $q^*$ of the groups correspond very well to the real groups, and the last group contains 
  only a very small fraction of nodes. 
  Depending on the initialization, this last group can even contain no node at all, in which 
  case it can be simply discarded. 
  This phenomenon is likely to come from the stochasticity of the SBM and is observed also for 
  large networks with $10^5$ nodes. 
  The modularity of those partitions is equal to, or slightly higher than those found in the $\hat{q}=q^*$ phase 
  of \modbp run with $q=q^*$, which explains why they are found. On the other hand, we have never observed more than one 
  of these additional, and almost empty, groups, such that $\hat{q}$ is always at most equal to $q^*+1$.  
  
  The other phenomenon 
  is that of distinct groups merging together
  in the retrieval partition, leading to $\hat{q}<q^*$. Such partitions have lower modularities than partitions with 
  $\hat{q}=q^*$ (found at same temperature from a different initialization), showing that the algorithm is unable to 
  correctly maximize the modularity starting from any initialization.
  This is likely due to the existence of ``hard but detectable'' phases~\cite{decelle2011asymptotic}, 
  in which frozen variables cause algorithms to be stuck in suboptimal solutions. A simple way out from this problem is to run the algorithm several times with different initial conditions, selecting finally the configuration of higher modularity found.
  
  These two effects might coexist, and produce retrieval partitions in which two of the real groups are merged into a single 
  one, and an additional group containing very few or even no nodes at all is also present.	
In this case $\hat{q}=q^*$, but the retrieval partition is not the right one.
So the existence of of an almost empty group should be considered as a warning on the reliability of the \modbp result.


\subsection{Results on real networks}

For community detection on real networks, $q^*$ is in general unknown and there is no available ground truth. 
From Fig.~\ref{fig:4} and the previous section, we know that \modbp can converge to partitions with different $\hat{q}$
at the same temperature, depending on the initialization. This motivates us to run \modbp several times for each temperature, 
which allows us to quantify the probability a given $\hat{q}$ is found at any given temperature $T$.
 Fig.~\ref{fig:7} shows the coexistence of phases in the ``political books''~\cite{polbooks} and ``political blogs''~\cite{adamic2005political} datasets for different values of $q$.
The analysis made in these figures is similar to the one proposed in~\cite{ronhovde2009multiresolution} for multiresolution community detection.

\begin{figure}
 \centering
 \includegraphics[width=\columnwidth]{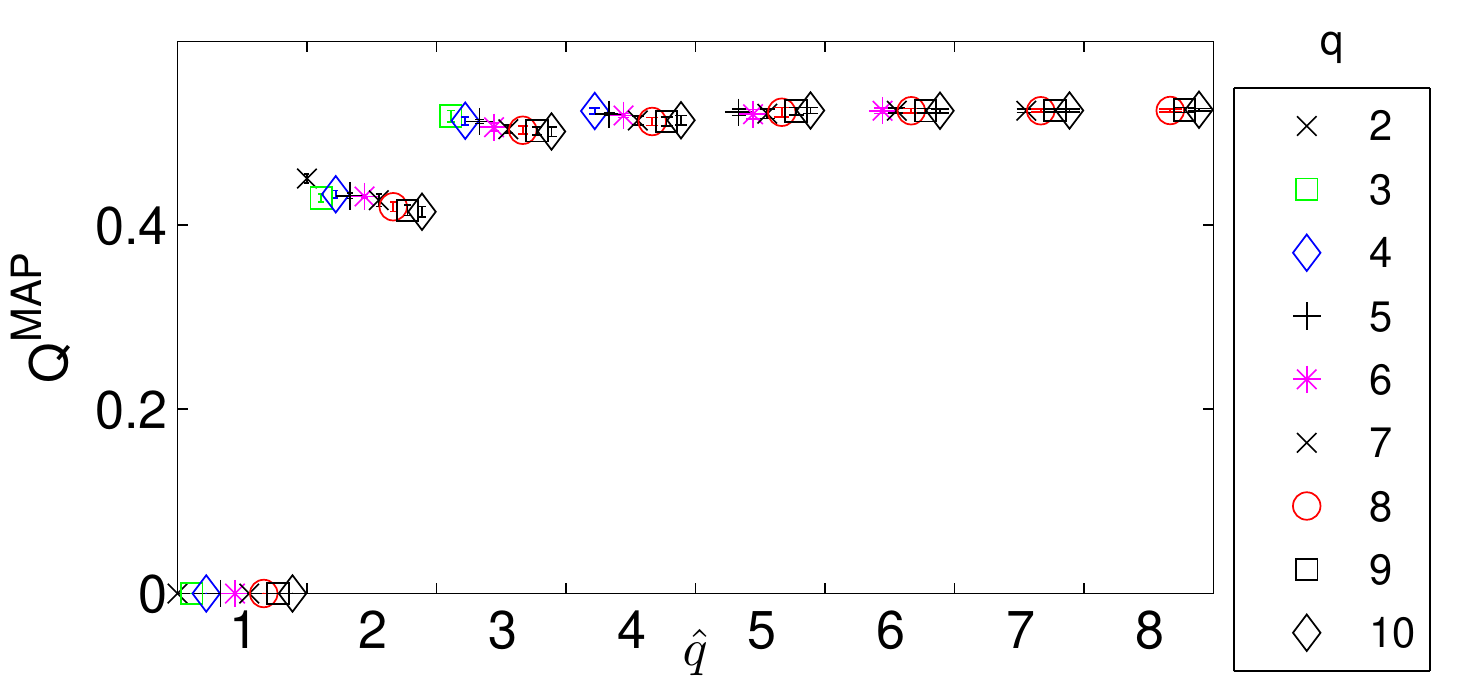}
 \caption{(Color online) $Q^{\rm MAP}$ as a function of $q$ and $\hat{q}$, using the same experimental results as in Fig.~\ref{fig:7}.
 Symbols represent the mean $Q^{\rm MAP}$ of all experiments with a given $q$ resulting in a given $\hat{q}$, along with 
 an error bar representing the standard deviation. Despite the use of different temperatures, the error standard deviations 
 are very small for each $q$. Furthermore, the mean $Q^{\rm MAP}$ for different $q$ are very similar, such that we can consider 
 $Q^{\rm MAP}$ to essentially depend on $\hat{q}$, with only negligeable influence of $q$ and $T$. 
 The fact that the increase in $Q^{\rm MAP}$ for $\hat{q}>3$ is minimal concords with the fact that $q^*=3$. }
 \label{fig:8}
\end{figure}

\begin{figure}
 \centering
 \includegraphics[width=\columnwidth]{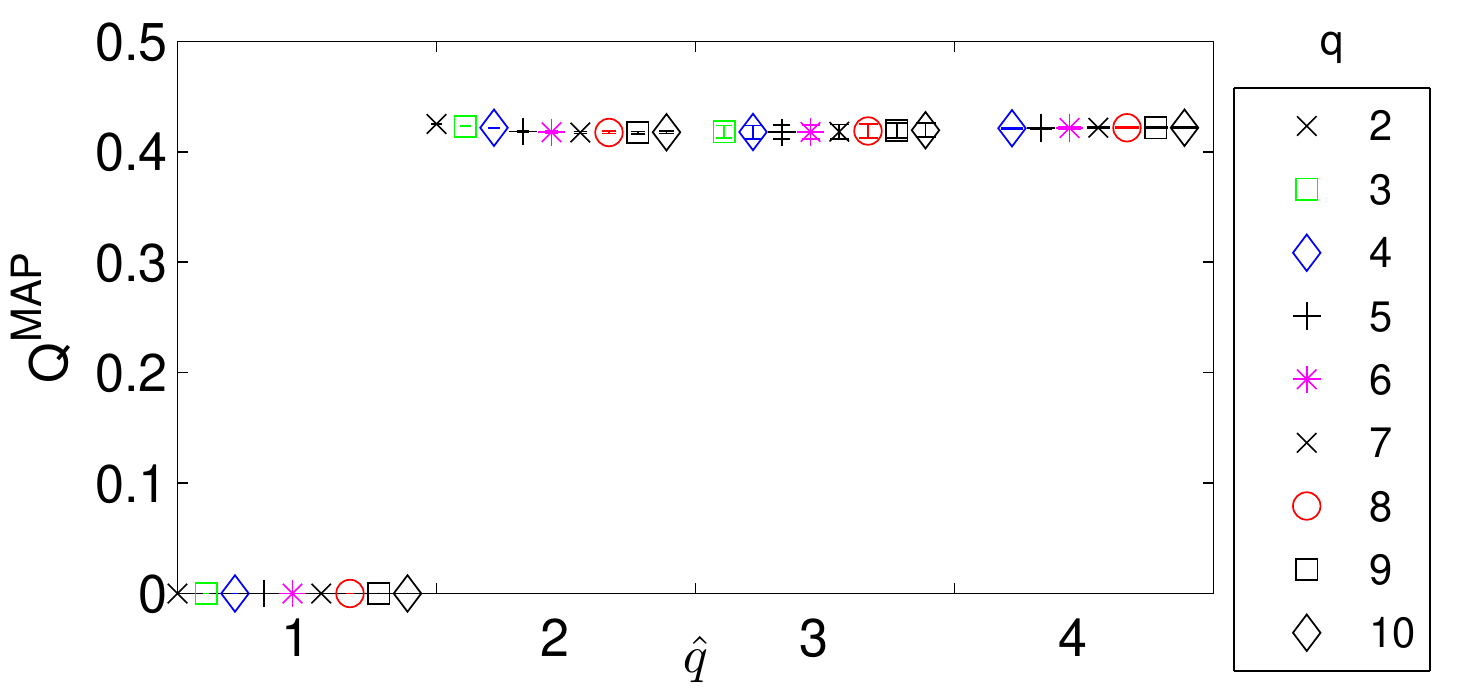}
 \includegraphics[width=\columnwidth]{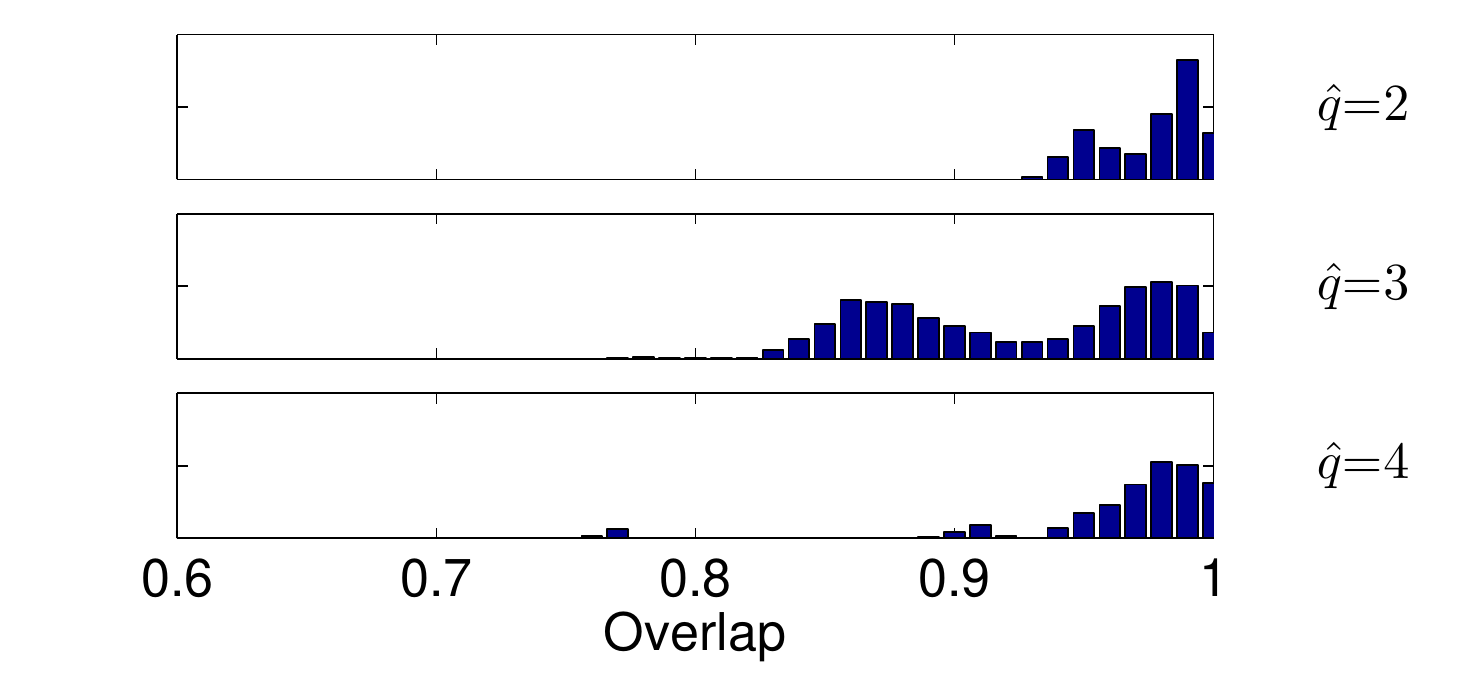}
 \caption{(Color online) \textbf{Top}:$Q^{\rm MAP}$ as a function of $q$ and $\hat{q}$, using the same experimental results as in Fig.~\ref{fig:7}.
 Symbols represent the mean $Q^{\rm MAP}$ of all experiments with a given $q$ resulting in a given $\hat{q}$, along with 
 an error bar representing the standard deviation. 
 The fact that $Q^{\rm MAP}$  does not increase for $\hat{q}>2$ concords with the fact that $q^*=2$. 
 \textbf{Bottom}: Empirical distribution of $20000$ overlaps between pairs of partitions with same $\hat{q}$.
 }
 \label{fig:8bis}
\end{figure}

These figures suggest that, at a given normalized temperature $T/T_0$,
the results returned by \modbp only marginally depend on the chosen $q$ as 
long as $q>q^*$. Moreover we observe that, within a phase with a given number $\hat{q}$ of groups found, the partition $\{t\}$ only marginally depends on the temperature $T$. Averaging over the several partitions found at different temperatures and with different initial condition, we show in Fig.~\ref{fig:8} and~\ref{fig:8bis} that $Q^{\rm MAP}$ depends essentially on $\hat{q}$, and only minimally on $q$.  As in ZM, we consider that the largest $\hat{q}$ leading to a significant increase of $Q^{\rm MAP}$ w.r.t.\ $\hat{q}-1$ is a plausible estimate of $q^*$, which agrees well 
with the commonly accepted ``ground truths'' of $q^*=3$ for ``political books'' and $q^*=2$ for ``political blogs''.

In Fig.~\ref{fig:8bis} we also show the distribution of overlaps between randomly chosen partitions with the same $\hat{q}$.
The overlap between two partitions $\{t\}$ and $\{s\}$ is a number between zero and one and is defined as
\begin{equation}
 O(\{t\},\{s\}) = \max_{\sigma} \left( \frac{1}{N} \sum_{i=1}^N \delta_{t_i, \sigma(s_i)} \right),
\end{equation}
where the maximum over all permutations $\sigma$ of $\{1, \hdots, \hat{q}\}$ allows to lift the permutation symmetry of the group 
numbering choice.
The closer the overlap between two partitions is to one, the more similar they are.
Fig.~\ref{fig:8bis} thus shows that partitions with the same $\hat{q}$ are very similar to one another, independently of the two 
parameters of \modbp, $T$ and $q$, for which they were obtained.
One may be worried about the double peak structure of the $\hat{q}=3$ case and wondering whether the two peaks 
do actually corresponds to different communities structures.
We have looked at the groups partitions returned by the algorithm and found the following. 
There is always a well conserved group of 520 to 530 nodes, while the remaining roughly 700 nodes can be clustered in different ways:
for $\hat{q}=3$, there are 2 different partitions with roughly 600+100 and 500+200 nodes;
for $\hat{q}=4$, the partition is roughly 380+280+40 nodes.
All these configurations have essentially the same modularity. 
So the conclusion is that the $\hat{q}=2$ partition (520+700 nodes) is significant, while further splitting of the cluster of 700 nodes is not very meaningful.

 

\section{Discussion}
Apart from the advantage of not requiring the knowledge of the generative model, a futher advantage of \modbp is that it has only two adjustable parameters, $T$ and $q$. 
However, for a given network, it is not clear how to choose them in order to obtain the optimal partition.
The recommendation of ZM is to run \modbp at $T^*(q)$, defined in Eq.~(\ref{eq:tstar}), for increasing values of $q$,
until it does not lead anymore to a significant increase in modularity.
Based on the experiments on synthetic and real networks presented in this work, we conclude that an 
important additional step in this procedure is to calculate the effective number of groups $\hat{q}$ of each partition 
returned by the algorithm, which can be different from $q$. Furthermore, this phenomenon leads to a new rule for 
assigning a group to each node, given that some groups might be merged, which also affects the modularity.

Another possible way to proceed is to run \modbp with a large value of $q$, and sweep the temperature scale from $T_0(q)$ downwards. As $T$ is lowered, the network is clustered into an increasing number of effective groups $\hat{q}$, and the found partitions have increasing modularities. Again, the procedure can be stopped once the modularity does not increase anymore in a significant way as $\hat{q}$ is increased.
 
For real networks, where the generating process is in general not known and not as straightforward as in the SBM, the number of groups to cluster the nodes is in part let as a choice to the user.
In this case, running \modbp with a quite large value of $q$ and using $T$ as the parameter to search for the optimal partition seem both desirable and efficient.
To make the optimal choice, in addition to the value of the modularity of a partition with $\hat{q}$ groups, the range of temperatures where this $\hat{q}$ phase exists might indicate how relevant it is (as shown in Fig.~\ref{fig:7}).
In particular, if a $\hat{q}$ phase only exists on a narrow range of temperatures, then it is likely to be less important, because less stable with respect to changes in the model parameter ($T$ in the present case).
 
Furthermore, as seen on graphs generated by the SBM, it may occur that some group contains a very small number of nodes.
In this case, merging them with bigger groups will only slightly change the modularity and result into a more meaningful and stable partition.

\section{Conclusion}
In this paper, we have studied the \modbp algorithm proposed in Ref.~\cite{zhang2014scalable}, focussing
on the influence of the choice of the two adjustable parameters $q$ and $T$, on both real and synthetic networks.
We have given a more precise picture of the algorithm behaviour by identifying new order parameters that allow 
to define several different phases inside the recovery phase. In each of these phases, \modbp clusters the nodes into 
a different number of groups $\hat{q}$. These phases can either be well separated on the temperature scale and 
be accompanied by a divergence in the number of iterations of the algorithm, or coexist on in the low temperature regime.
The partitions with the same number $\hat{q}$ of groups typically have high overlaps among them and very similar modularities.
We have proposed a normalized temperature scale ($T/T_0$) on which \modbp has a very similar behavior for different values of $q$.
These findings provide a broader description of the \modbp algorithm behaviour, showing its robustness and effectiveness.
Hopefully they can be very useful when \modbp is run on real networks where the ground through is unknown.

Real network may have hierarchical structures~\cite{girvan2002community,ronhovde2009multiresolution,peixoto2014hierarchical,nussinov2015inference}  and the deeper understanding of the different recovery phases achieved
in this work may help in using the temperature as a simple parameter to study by \modbp different levels of the hierarchical clustering. 
The different levels of clustering hierarchy may correspond to recovery phases with different values of $\hat{q}$, obtained decreasing the temperature.

\section*{Acknowledgement}
The authors thank Cristopher Moore, Lenka Zdeborova and Pan Zhang for useful discussions. This research has received funding from the Italian Research Ministry through the FIRB Project No. RBFR086NN1.
C.S. is funded by the Universit\'e franco-italienne.

\bibliography{refs}

\end{document}